\documentclass[prd,twocolumn,twoside,preprintnumbers,superscriptaddress,nofootinbib,natbib,bm,amsrefs]{revtex4-2}
\usepackage{graphicx}
\usepackage{hyperref}
\usepackage{url}

\usepackage{xcolor}
\usepackage{amsmath, amstext, amsfonts,amssymb, amsthm}

\newcommand{\Eprint}[1]{\href{#1}}

\definecolor{lb}{rgb}{.74,.83,.9}
\definecolor{ly}{rgb}{1,.92,.8}
\definecolor{lr}{rgb}{.98,.85,.87}

\begin{document}

\preprint{PSI-PR-23-7, ZU-TH 15/23, P3H-23-015, TTP23-010, KEK-TH-2506}

\title{Asymmetric di-Higgs signals of the next-to-minimal 2HDM with a \texorpdfstring{\boldmath{$U(1)$}}{U(1)} symmetry.}

\author{Sumit Banik}
\email{sumit.banik@psi.ch}
\affiliation{Physik-Institut, Universität Z\"urich, Winterthurerstrasse 190, CH–8057 Zürich, Switzerland}
\affiliation{Paul Scherrer Institut, CH–5232 Villigen PSI, Switzerland}

\author{Andreas Crivellin}
\email{andreas.crivellin@cern.ch}
\affiliation{Physik-Institut, Universität Z\"urich, Winterthurerstrasse 190, CH–8057 Zürich, Switzerland}
\affiliation{Paul Scherrer Institut, CH–5232 Villigen PSI, Switzerland}

\author{Syuhei Iguro}
\email{igurosyuhei@gmail.com}
\affiliation{Institute for Theoretical Particle Physics (TTP), Karlsruhe Institute of Technology (KIT), Engesserstra{\ss}e 7, 76131 Karlsruhe, Germany}
\affiliation{Institute for Astroparticle Physics (IAP), KIT, Hermann-von-Helmholtz-Platz 1, 76344 Eggenstein-Leopoldshafen, Germany}

\author{Teppei Kitahara}
\email{teppeik@kmi.nagoya-u.ac.jp}
\affiliation{Institute for Advanced Research \& Kobayashi-Maskawa Institute for the Origin of Particles and the Universe, 
Nagoya University,  Nagoya 464--8602, Japan}
\affiliation{KEK Theory Center, IPNS, KEK, Tsukuba  305--0801, Japan}
\affiliation{CAS Key Laboratory of Theoretical Physics, Institute of Theoretical Physics, Chinese Academy of Sciences, Beijing 100190, China}

\begin{abstract}
The two-Higgs-doublet model with a $U(1)_H$ gauge symmetry (N2HDM-$U(1)$) has several advantages compared to the ``standard'' $Z_2$ version (N2HDM-$Z_2$): It is purely based on gauge symmetries, involves only spontaneous symmetry breaking, and is more predictive because it contains {one parameter less} in the Higgs potential, which further ensures $CP$ conservation, {i.e., avoiding the stringent bounds from electric dipole moments}.
After pointing out that a second, so far unknown version of the N2HDM-$U(1)$ exists, we examine the phenomenological consequences {for the Large Hadron Collider} (LHC) of the differences in the scalar potentials. In particular, we find that while the  N2HDM-$Z_2$ predicts suppressed branching ratios for decays into different Higgs bosons for the case of the small scalar mixing (as suggested by Higgs coupling measurements), both versions of the N2HDM-$U(1)$ allow for sizable rates. This is particularly {relevant} in light of the CMS excess in  {resonant} Higgs-pair production at around $650\,$GeV  {of a} Standard Model Higgs boson {subsequently} decaying to photons and a new scalar with a mass of $\approx90\,$GeV  {subsequently} decaying to bottom quarks (i.e., compatible with the CMS and ATLAS $\gamma\gamma$ excesses at $95\,$GeV and $\approx 670\,$GeV). {As we will show, this excess can be  {addressed} within the N2HDM-$U(1)$ {in case of a nonminimal Yukawa sector}, predicting an interesting and unavoidable $Z+ b\bar b$ {signal} and motivating further asymmetric di-Higgs searches at the LHC.}
\end{abstract}
\maketitle

\section{Introduction}
The discovery of the Brout-Englert-Higgs boson~\cite{Higgs:1964ia,Englert:1964et,Higgs:1964pj,Higgs:1964pj,Guralnik:1964eu,Higgs:1966ev,Kibble:1967sv} by ATLAS~\cite{Aad:2012tfa} and CMS~\cite{Chatrchyan:2012ufa} established, for the first time, the existence of a fundamental scalar particle within the Standard Model (SM). This observation motivates the existence of more scalar particles and, in turn, the experimental search for them. While the $125\,$GeV Higgs ($h$) boson has approximately SM-like properties~\cite{Khachatryan:2014kca,Aad:2015mxa,Khachatryan:2014jba,Aad:2015gba,CMS:2022dwd,ATLAS:2022vkf}, this does not exclude the existence of additional scalar bosons, as long as their role in the breaking of the electroweak (EW) gauge symmetry and the mixing with the SM-like Higgs boson are sufficiently small. 

In this context, strong constraints on new physics (NP) models are provided by the $\rho$ parameter that relates the EW gauge couplings to the $W$ and $Z$ masses and is defined to be unity in the SM at tree level. This singles out models with $SU(2)_L$-singlet or $SU(2)_L$-doublet extensions of the SM Higgs sector whose vacuum expectation values (VEVs) conserve the custodial symmetry, such that the additional scalars only give loop-level effects in the $\rho$ parameter.\footnote{Also larger $SU(2)_L$ multiplets are possible in case their VEV is very small, or if a global custodial  $SU(2)$ symmetry protects the $\rho$ parameter, which however entails larger and more complex Higgs sectors~\cite{Chanowitz:1985ug}.}

The most studied extensions of the SM scalar sector are the two-Higgs-doublet models (2HDMs)~\cite{Lee:1973iz,Gunion:1986nh,Gunion:1989we,Branco:2011iw}. Here, usually a discrete $Z_2$ symmetry is imposed to both solve the problem of flavor changing neutral currents~\cite{Glashow:1976nt,Paschos:1976ay} (resulting in four different versions with natural flavor conservation~\cite{Barger:1989fj,Aoki:2009ha}) and to provide (accidental) $CP$ conservation in the Higgs potential. However, for phenomenological reasons, i.e., to give VEV-independent masses to the additional scalars, the $Z_2$ symmetry must be broken. In order to avoid domain wall problems caused by a spontaneous discrete symmetry breaking~\cite{Zeldovich:1974uw}, the $Z_2$ symmetry is usually softly broken by a dimension-two term~\cite{Battye:2020jeu}. This operator (in case of a nonvanishing and non-aligned $\lambda_5$ term) in general gives rise to $CP$ violation within the Higgs sector, with potentially dangerously significant effects in electric dipole moments~\cite{Chupp:2017rkp}. 

Reference~\cite{Ko:2012hd} proposed to solve these problems by replacing the discrete $Z_2$ symmetry with a $U(1)_H$ gauge symmetry, which can mimic the effect of the $Z_2$ symmetry but forbids the explicit soft-breaking term. However, if the $Z^\prime$ boson originating from the  $U(1)_H$ gauge is required to be heavier than the EW scale, one has to supplement the model with an additional scalar charged under $U(1)_H$; minimally a complex scalar $\phi$ that is a singlet under the SM gauge group. Because its $CP$-odd component becomes (in the vanishing scalar mixing limit) the longitudinal component of the $Z^\prime$, the scalar potential effectively resembles the one of the Next-to-Minimal 2HDM (N2HDM) with a real scalar (see, e.g., Refs.~\cite{He:2008qm,Grzadkowski:2009iz,Logan:2010nw,Boucenna:2011hy,He:2011gc,Bai:2012nv,He:2013suk,Cai:2013zga,Chen:2013jvg,Guo:2014bha,Wang:2014elb,Drozd:2014yla,Campbell:2015fra,Drozd:2015gda,vonBuddenbrock:2016rmr,Arhrib:2018qmw,Engeln:2020fld,Azevedo:2021ylf,Biekotter:2022abc}). In particular, the VEV of $\phi$ gives rise to the $m_{12}^2$ term that softly breaks the $Z_2$ symmetry.

While in the N2HDM with two discrete $Z_2$ symmetries (N2HDM-$Z_2$)\footnote{Under the first $Z_2$ symmetry only $H_2$ is odd, while for the second $Z_2$ only the real scalar is odd.}, 
Higgs-boson-related collider observables have been studied in detail~\cite{Abouabid:2021yvw}, including loop effects~\cite{Drechsel:2016htw,Baglio:2019nlc}, and even automated codes exist~\cite{Engeln:2018mbg,Krause:2019oar,Muhlleitner:2020wwk,Bahl:2022igd},\footnote{The program ewN2HDECAY \cite{Krause:2019oar} is based on HDECAY~\cite{Djouadi:1997yw,Djouadi:2018xqq} and 2HDECY~\cite{Krause:2018wmo}.} studies of the N2HDM with a $U(1)_H$ gauge symmetry (N2HDM-$U(1)$) have not focused on the collider phenomenology of the additional scalars but mostly considered dark matter~\cite{Ko:2014uka,Camargo:2019ukv,Nomura:2019wlo}, muon $g-2$\,\cite{Nomura:2020kcw}, neutrino masses~\cite{Ko:2014tca,Nomura:2017wxf,Bertuzzo:2018ftf,Camargo:2018uzw,Cai:2018upp,Dias:2021lmf}, $b\to s\ell^+\ell^-$ anomalies~\cite{Crivellin:2015lwa,Bian:2017rpg}, $Z^\prime$ searches~\cite{Camargo:2018klg,Aguilar-Saavedra:2022rvy,Aad:2012tfa}, and Higgs signal strengths~\cite{Ko:2013zsa}. However, also the (effective) scalar potential of this N2HDM-$U(1)$ is different from the one of the usual N2HDM-$Z_2$, which, in particular, leads to different Higgs self-interactions and, therefore, different decay rates for heavy scalars into light ones.

This aspect is now more relevant in light of the ongoing and intensified Large Hadron Collider (LHC) searches for new scalar bosons (see, e.g., Ref.~\cite{Naryshkin:2021ryt} for a recent review). While no unequivocal evidence for a new particle has been observed, interesting hints for new scalars with masses around 95$\,$GeV~\cite{LEPWorkingGroupforHiggsbosonsearches:2003ing,CMS:2018cyk,CMS:2022rbd,CMS:2022tgk,Cao:2016uwt,Crivellin:2017upt,Haisch:2017gql,Fox:2017uwr,Heinemeyer:2018jcd,Biekotter:2019kde,Cao:2019ofo,Biekotter:2022jyr,Iguro:2022dok,Biekotter:2022abc,Iguro:2022fel,Coloretti:2023wng}, 151$\,$GeV~\cite{vonBuddenbrock:2017gvy,ATLAS:2021jbf,Crivellin:2021ubm,Richard:2021ovc,Fowlie:2021ldv} and 670$\,$GeV~\cite{CMS:2017dib,ATLAS:2021uiz,CMS:2022tgk,Consoli:2021yjc} have been reported.\footnote{In addition, anomalies in multilepton final states exist, which can also be explained by new scalars~\cite{vonBuddenbrock:2017gvy,vonBuddenbrock:2019ajh,vonBuddenbrock:2020ter,Hernandez:2019geu,Fischer:2021sqw}.} In particular, the CMS excess for a $\approx650\,$GeV scalar decaying into a $\approx90\,$GeV one (i.e., compatible with the 95$\,$GeV hints mentioned above due to the limited detector resolution for bottom jets) and a SM Higgs~\cite{CMS:2022tgk} needs {multiple} new Higgs bosons with a mass hierarchy (which is only possible if they are not within the same $SU(2)_L$  {multiplet while} respecting perturbativity bounds)\footnote{Note that the di-Higgs excess is compatible with the 95$\,$GeV and 670$\,$GeV hints mentioned before due to the limited detector resolution for bottom jets, which effectively removes the look-elsewhere effect.}\footnote{ {The mass differences between the additional scalars are controlled by the couplings in the scalar potential if they come from the same multiplet. The large mass difference compared to the EW VEV requires the huge coupling, and hence violates the unitarity constraint.}}. Furthermore, as it is an asymmetric di-Higgs decay, it requires {interactions among} the three different scalars. Here, asymmetric decay means
the decay of a heavy scalar into two {different} lighter scalars.
While the rates of such asymmetric di-Higgs decays are in general small in the MSSM~\cite{Plehn:1996wb,Dawson:1998py}, 2HDMs~\cite{Baglio:2023euv} and also in the N2HDM-$Z_2$~\cite{Muhlleitner:2016mzt}, we will show that for the N2HDM-$U(1)$ they are naturally sizable.

\section{The Model}

As outlined in the Introduction, a $Z_2$ symmetry is commonly used to construct the four versions of the 2HDMs with natural flavor conservation and, at the same time, constrain the scalar potential. In the N2HDM, even two $Z_2$ symmetries are usually employed to prevent tree level flavor changing neutral currents and eliminate most sources of $CP$ violation. We want to use instead a single $U(1)_H$ gauge symmetry under which at least two of the scalar fields are charged. 

We start with the scalar potential for the two $SU(2)_L$ doublets $H_1$ and $H_2$ with hypercharge $1/2$ (where according to the usual 2HDM conventions $H_2$ contains most of the SM Higgs {for the case of small mixing angles}). If the $U(1)_H$ charges of $H_1$ and $H_2$ are different, operators with an odd number of these fields are forbidden, leading to
\begin{align}
    {\cal V}_{H} =& m_{11}^2|H_1|^2 + m_{22}^2|H_2|^2 + \frac{\lambda _1}{2}(H_1^\dag {H_1})^2 + \frac{\lambda _2}2(H_2^\dag {H_2})^2 \nonumber\\&+ \lambda _3(H_1^\dag H_1)(H_2^\dag H_2) + \lambda_4(H_1^\dag H_2)(H_2^\dag H_1)\,.
\end{align}
This potential is $CP$ conserving as it does not contain the soft-breaking term $m_{12}^2 H_1^\dag H_2$ nor the term $\frac{\lambda _5}{2}(H_1^\dag {H_2})^2$ contained in the 2HDM with the (softly broken) $Z_2$ symmetry. 

Next, we add a complex scalar SM singlet $\phi$ that is charged under the $U(1)_H$ gauge symmetry. Its self-interactions, as well as the ones with two identical doublets
\begin{align}
{\cal V}_{\phi} &= |\phi|^2\left( m_\phi^2 + \frac{{{\lambda _\phi}}}{2}|\phi|^2 + {\lambda _{\phi1}}|H_1|^2 + {\lambda _{\phi2}}|H_2|^2\right)\,,
\end{align}
are allowed independently of the $U(1)_H$ charges. In addition, there are two options for charge assignments under the $U(1)_H$ symmetry {(for $Q_H(H_1) \neq Q_H(H_2)$)}:
\newline
{\bf (a)} If $|Q_H(\phi)|=|Q_H(H_1)-Q_H(H_2)|$, one has the term
    \begin{align}
{\cal V}_{\phi H}^{\rm a}= \sqrt{2}\mu H_1^\dag H_2\phi + {\rm h.c.} \,,
\label{casea}
\end{align}
or $\phi$ replaced by $\phi^\dagger$, depending on the sign on the $U(1)_H$ charge.
\newline
{\bf (b)} If $|Q_H(\phi)|=|Q_H(H_1)-Q_H(H_2)|/2$, the term
\begin{align}
{\cal V}_{\phi H}^{\rm b}={\lambda_{\phi12}}(H_1^\dag {H_2}) \phi^2 + {\rm{h}}{\rm{.c}}{\rm{.}}\,,
\label{caseb}
\end{align}
is gauge invariant. Case~(a) was already proposed in Ref.~\cite{Ko:2012hd}, while case~(b) is novel, to the best of our knowledge. 

Note that we have normalized the prefactors of these potentials in such a way, that once we decompose
\begin{equation}
    \phi=(v_S + \hat S+i\eta_S)/\sqrt{2}\,,
\end{equation}
$\eta_S$ (mostly) becomes the longitudinal mode of the $Z^\prime$ and the terms involving $\hat S$ match the N2HDM-$Z_2$. Here, $v_S$ is the VEV of $\phi$, and one can choose it to be real and positive without loss of generality. Therefore, disregarding the $Z^\prime$ boson, which could be heavy or weakly coupled, the N2HDM-$U(1)$ resembles the N2HDM-$Z_2$ with the important differences that the $m_{12}^2$ and $\lambda_5$ terms are only effectively generated by $v_S$ and $Z^\prime$ exchange, respectively, similar to the $\mu$ term in the next-to-minimal supersymmetric Standard Model (NMSSM)~\cite{Fayet:1974pd,Barbieri:1982eh,Dine:1981rt,Nilles:1982dy,Frere:1983ag,Derendinger:1983bz}.  {Moreover}, this  {guarantees} the absence of $CP$ violation in the scalar potential (even when the $Z^\prime$ is integrated out), while this, in general,  {is inevitable} in both the N2HDM-$Z_2$ and NMSSM.

We know from Higgs signal strength measurements that the mixing among the SM-like Higgs boson and the other two $CP$-even scalars should be rather small. Therefore, we will label the $CP$-even mass eigenstates, contained in the absence of mixing within $H_2$, $H_1$, and $\phi$ as $h$, $H$, and $S$, respectively.\footnote{As we only consider the case of small mixing, we will label in the main text both the $CP$-even components of the doublets and the singlet, as well as the mass eigenstates by $h$, $H$, and $S$ and use them interchangeably. In Appendix~\ref{AppA}, the full mass matrices in the weak eigenbasis are given.} 
Importantly, the mixing {among} $H$, $h$ and $S$ in the N2HDM-$U(1)$ is related to the masses $m_H$, $m_{H^\pm}$, and $m_{A}$ (where $H^\pm$ and $A$ denote the charged and $CP$-odd Higgs boson, respectively) because they all involve the effective $m_{12}^2$ term originating from $\mu$ or $\lambda_{\phi12}$. This means that the effective $m_{12}^2$ term automatically leads to $H$-$S$ mixing as can be inferred from the $CP$-even mass matrix {(in the large $\tan\beta$ limit, {i.e.,~assuming $v_1$ to be small})} 
\begin{equation}
 M^2_\rho \approx \begin{pmatrix}
  - \mu v_S\tan\beta & \mu v_S &
   \mu v \\  \mu v_S & \lambda_2 v^2 &
   \lambda_{\phi2} v v_S \\
 \mu v & \lambda _{\phi2} v v_S & \lambda _\phi v_S^2  \\
\end{pmatrix}\,,
\label{massmatrix}
\end{equation}
where $\tan\beta=v_2/v_1$ and $\langle H_i\rangle = v_i/\sqrt{2}$. Note that the effects of $\lambda_1, \lambda_{3}, \lambda_{4}$ and $ \lambda_{\phi 1}$ on the {$CP$-even Higgs} masses become negligible in the large $\tan\beta$ {limit}. The full expressions for the minimization, the mass matrices, etc., can be found in Appendix~\ref{AppA}.

Concerning the fermion Yukawa sector, the most natural choice is probably to assume that {the SM fermions} are uncharged under $U(1)_H$, or to assign equal charges to left-handed and right-handed fields (such as $B-L$ or $L_\mu-L_\tau$) in order to avoid gauge anomalies. In this setup, the doublet $H_2$ would be $U(1)_H$ neutral, while $H_1$ carries some $U(1)_H$ charge $Q_H$. This then leads to a type-I Yukawa sector, which also has the advantage of being quite unconstrained in the large $\tan\beta$ and small $\alpha$ (Higgs mixing angle) limit. However, also the other three types of 2HDMs with natural flavor conservation, as well as the generic type-III model,\footnote{The type-III model has been comprehensively studied in Ref.~\cite{Crivellin:2013wna} as it can (partially) explain the tensions in $R(D^{(*)})$~\cite{Crivellin:2012ye,Blanke:2022pjy}.} can be obtained even in an anomaly free fashion if the fermion sector is extended~\cite{Ko:2012hd}. 

\section{Phenomenology}

The N2HDM-$U(1)$ is in general more predictive than the N2HDM-$Z_2$ as it contains {one parameter less} and has no sources of $CP$ violation in the Higgs potential. However, what is the most striking difference regarding LHC observables between the different N2HDMs, even when the $Z^\prime$ predicted by the N2HDM-$U(1)$ is disregarded, as it might be heavy or weakly coupled? 

To answer this, let us consider the limit of vanishing mixing among the neutral $CP$-even scalars, in which $h$ is purely SM-like, $H$ only couples to $W^\pm H^\mp$ and $ZA$, and $S$ is sterile. Now, $\mu$ ($\lambda_{\phi12}$) in Eq.~(\ref{casea}) (Eq.~(\ref{caseb})) has to be nonvanishing to give masses to $H$, $A$, and $H^\pm$ that are above the EW scale, i.e.,
\begin{equation}
    m_H^2\approx m_A^2 \approx m_{H^\pm}^2 \approx - \mu v_S \tan\beta\,,
\end{equation}
{assuming $\lambda_4$ to be small}.
This then, at the same time, {suppresses} $H$-$h$ and $H$-$S$ mixing{s}, and from Eq.~(\ref{massmatrix}) we see that $h$-$S$ mixing can be avoided, {for large $\tan\beta$ with} $\lambda_{\phi2}=0$. This means that for $m_H\gg v$ the only unsuppressed decay of $H$, in the absence of Yukawa coupling of $H_1$, is $H\to Sh$ for case (a), and in addition $H\to SS$ for case of (b), if $m_H\gg m_S$ and $m_{H^\pm}\approx m_H$. Therefore, in the large $\tan\beta$ limit, the N2HDM-$U(1)$ predicts sizable branching ratios for $H\to Sh$ (and also $H\to SS$ in case (b)). As this decay in the N2HDM-$Z_2$ is suppressed by small mixing angles, the discovery of an asymmetric di-Higgs signal would be a smoking gun for the N2HDM-$U(1)$.\footnote{This resembles the situation in the NMSSM where also sizable asymmetric Higgs decays are possible~\cite{NMSSM-benchmark,Ellwanger:2022jtd}.}

\begin{figure}
    \centering
    \includegraphics[scale=0.5]{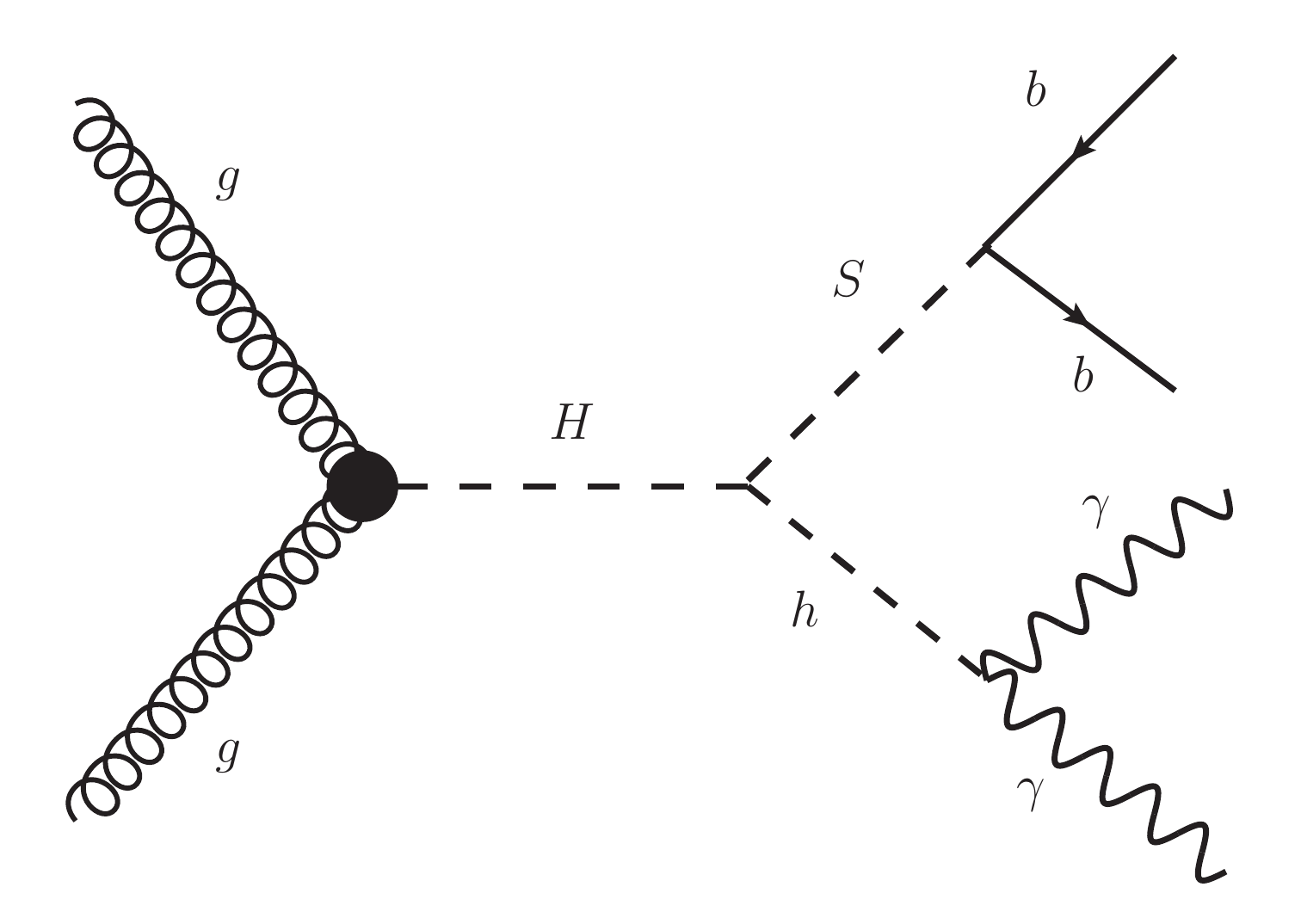}
    \caption{Feynman diagram showing resonant asymmetric Higgs pair production. The discovery of this process, for which the CMS measurement constitutes a first hint, would be a smoking gun for the N2HMD-$U(1)$. Here, the black circle denotes the loop induced effective coupling to gluons. However, note that the heavy top limit cannot be used because $m_t \ll m_H$ and we use the expression for a dynamical top quark in our numerical analysis.}
    \label{fig:my_label}
\end{figure}

Let us now illustrate this observation more quantitatively in the context of the hint for the $\approx650\,$GeV boson decaying into a $\approx90\,$GeV scalar and the SM Higgs boson with a global (local) significance of $2.8\,\sigma$ ($3.8\,\sigma$)~\cite{CMS:2022tgk}. Here, the $\approx90\,$GeV resonance decays into $b\bar b$ and the SM Higgs boson into $\gamma\gamma$. Because the detector resolution for bottom jets is not so good, this $\approx90\,$GeV excess could be compatible with the $\gamma\gamma$~\cite{CMS:2018cyk}, $\tau\bar\tau$~\cite{CMS:2022rbd} and the LEP $ZH$ measurement~\cite{LEPWorkingGroupforHiggsbosonsearches:2003ing} pointing towards a mass of $\approx95\,$GeV,
as well as with the $\gamma\gamma$~\cite{ATLAS:2021uiz} and $ZZ$ excesses~\cite{CMS:2017dib} at around $\approx670\,$GeV. This makes this asymmetric di-Higgs signal particularly interesting and effectively eliminates the look-elsewhere effect of the CMS di-Higgs analysis.

The {cross section for the $pp\to b\bar b \gamma\gamma$ excess is $\approx0.35^{+0.17}_{-0.13}$\,fb according to the CMS analysis~\cite{CMS:2022tgk}. However, the CMS analysis of $pp\to b\bar b \tau\bar\tau$~\cite{CMS:2021yci} finds an upper limit on the corresponding cross section of $\approx 4\,$fb {for $\approx 650\,$GeV boson search with $m_{bb}\approx 90\,$GeV}. With ${\rm BR}(h\to \tau\bar\tau)/{\rm BR}(h\to \gamma\gamma)\approx20$, {one can obtain $\approx 0.2\,$fb as the upper bound on $ pp\to b\bar b \gamma\gamma$, and} this translates into the limit $\sigma(pp\to (650)\to (95)\, h)\times{\rm BR}((95)\to b\bar b) \lessapprox 90\,$fb,
taking into account that BR$(h\to\gamma\gamma)\approx0.23\%$.
Therefore, the $b\bar b \gamma\gamma$ excess cannot be fully explained, but it is still possible to account for it within $2\,\sigma$ and we will aim at
\begin{equation}
    \sigma (pp\to  (650)  \to  (95)\, h) \times {\rm BR}((95) \to  b\bar b) \approx70\,{\rm fb}\,.   \label{eq:650gevanomaly}
\end{equation}
}

There are two options within the N2HDM-$U(1)$;  {one can A) identify the $\approx95\,$GeV state with $H$ and the $\approx650\,$GeV one with $S$ ($pp\to S\to Hh$) or  B) vice versa ($pp\to H\to Sh$). However, in case A),} the $\mu$ term is naturally small because $H$ is light, such that also the branching ratio is suppressed, unless one chooses very small mixing angles among the $CP$-even scalars. Let us, therefore, consider {option B), i.e.,~$pp\to H\to Sh$} (shown in Fig. \ref{fig:my_label}) in the following, again in the limit of small mixing {and large $\tan\beta$.}
To obtain a sufficient production cross section of $H$ we will consider the case of a nonminimal flavor structure and {assume that $H$ has a (effective) coupling to top quarks originating from the Lagrangian term $-\tilde Y^{t} \bar Q_L \tilde H_1 t_R$.}\footnote{This coupling can be induced at tree level by vector like quarks mixing with SM ones via coupling to $S$. Alternatively, an effective coupling to gluons could be loop induced by colored new heavy fermions or scalars. In fact, CMS observed an excess in di-di-jet searches~\cite{CMS:2022usq} that point towards new colored particles at the TeV scale~\cite{Crivellin:2022nms}.} 
 This coupling then also leads to unsuppressed decays of $H\to t\bar t$ (and $A\to t\bar t$). 

For the numerical analysis we use that a SM Higgs boson with a mass of $\approx650\,$GeV would have a gluon fusion production cross section of $\approx1.35\,$pb at NNLO~\cite{LHCHiggsCrossSectionWorkingGroup:2016ypw,Spira:1995rr,Liebler:2016ceh,Graudenz:1992pv,Anastasiou:2006hc,Aglietti:2006tp}. This means that a coupling to top quarks is needed, that is around one quarter of the one of the SM Higgs boson, i.e., $\tilde Y_t\approx Y_t/4 /(\sqrt{{\rm BR}(H\to Sh)}\sqrt{{\rm BR}(S\to b\bar b)})$. Therefore, assuming that $S$ decays SM-like\footnote{ {This procedure is justified since the other couplings of $S$  are given by the mixing with the other scalars and hence aligned to the SM-like coupling structure.}} (BR$(S\to b\bar b)\approx 0.8$) results with Eq.~\eqref{eq:650gevanomaly} in $\sigma (pp\to H)\approx84\,{\rm fb}/{\rm BR}(H\to Sh)$. Based on the Goldstone boson equivalence theorem~\cite{Cornwall:1974km,Vayonakis:1976vz}, we also expect BR$(A\to SZ)\approx\text{BR}(H\to Sh)$ leading to $pp \to A \to Z S \to Z b\bar{b}$ (and also $pp \to A \to Z h \to Z b\bar{b}$)
with cross sections around $1.5 \times 70\,$fb,\footnote{Note that at this energy, the gluon-fusion-production cross section for a pseudoscalar via top-quark loops is $\approx 1.5$ times the one of a $CP$-even scalar with the same mass and coupling (see, e.g., Refs.~\cite{Dawson:1998py,Spira:2016ztx}).} searched for by ATLAS~\cite{ATLAS:2017otj,ATLAS:2020gxx,ATLAS:2022jsi} and CMS~\cite{CMS:2018amk,CMS:2019ogx}. Note that in fact, Ref.~\cite{CMS:2019ogx} finds a mild excess within the relevant region. 

\begin{figure}
    \centering
    \includegraphics[scale=0.5]{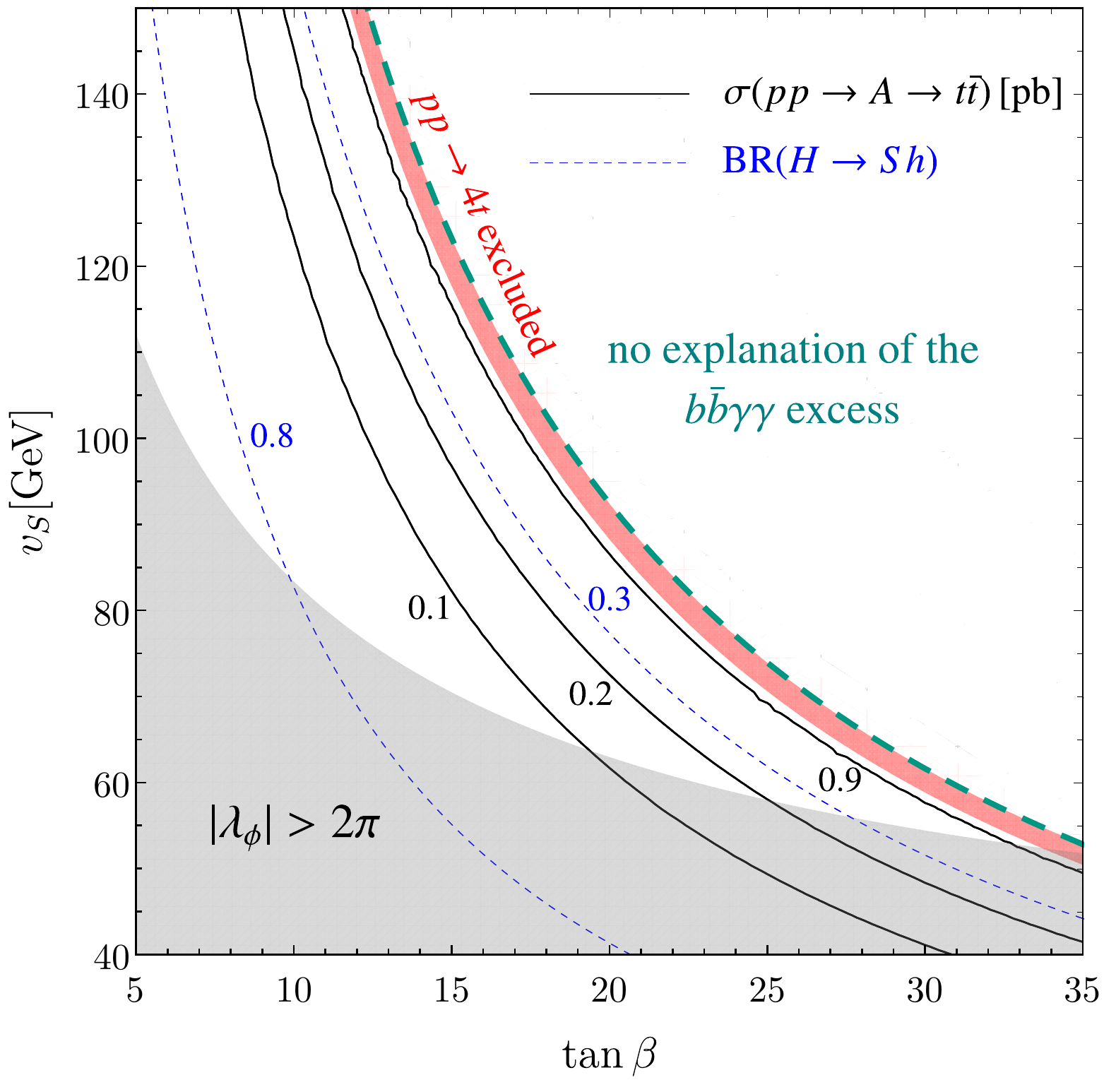}
    \vspace{-0.2cm}
    \caption{Predictions for $\sigma(pp\to A\to t\bar t)$[pb] as a function of $\tan\beta$ and $v_S$ in the N2HMD-$U(1)$ for case (a), assuming that the CMS excess $b\bar b \gamma\gamma $ in Eq.~\eqref{eq:650gevanomaly} is explained. The gray region is excluded by the requirement of perturbative couplings, while the red region is excluded by the $t\bar t t \bar t$ search \cite{ATLAS:2022rws}, assuming $m_A\approx m_H$. Note that the $b\bar b \gamma\gamma $ excess cannot be explained in the top-right region of the green dashed line.} 
    \label{fig:4t}
\end{figure}

Furthermore, we can predict the cross section of $H\to t\bar t$ and $A\to t\bar t$, as well as $pp\to t\bar t H\to t\bar t t\bar t$ and $pp\to t\bar t A\to t\bar t t\bar t$ as a function of $\tan\beta$ and $v_S$ (assuming $\lambda_{\phi2}=0$ as well as $m_H\approx m_A$) and compare this to the limits on the resonant $t\bar t$ production of CMS~\cite{CMS:2019pzc} and ATLAS~\cite{ATLAS:2020lks} as well as to $t\bar t t \bar t$ production measured by CMS~\cite{CMS:2019rvj} and ATLAS~\cite{ATLAS:2022rws}. This is illustrated in Fig.~\ref{fig:4t} where we show the predicted cross section for $pp\to A\to t\bar t$ in units of pb as a function of $\tan\beta$ and $v_S$. The red region is excluded by the $pp\to t\bar t t\bar t$ search of ATLAS and the  {gray} region by the requirement of positive eigenvalues of the mass matrix as well as perturbative couplings. Since $\sigma(pp\to H \to S h ) \approx 84\,$fb is required,
when BR$(S\to\gamma\gamma)\approx0.15\%$ (again assuming that $S$ has SM-like branching ratios) we obtain for the inclusive cross section $\sigma(pp\to S +X\to \gamma\gamma +X)\approx 0.1\,$fb which is compatible with the current limits but insufficient to explain the $\gamma\gamma$ excess at $95\,$GeV of $\approx 50\,$fb~\cite{CMS:2018cyk}. Therefore, direct production of $S$ would be required in addition to explain the $\gamma\gamma$ excess, {e.g.,~via gluon fusion from the mixing of $S$ with $H$ or $h$~\cite{Biekotter:2019kde}.}
{We would like to emphasize again that although the current excesses are not fully explained in the minimal model, the asymmetric di-Higgs decay is a key prediction and further experimental analyses are encouraged.}  

\section{Conclusions}

While Higgs physics in the N2HDM with two discrete $Z_2$ symmetries (N2HDM-$Z_2$) has been studied in detail in the literature, this phenomenological aspect of the N2HDM with a $U(1)_H$ gauge symmetry has received little to no attention so far. While both versions have desirable features such as natural flavor conservation, there are even several advantages of the N2HDM-$U(1)_H$ over the N2HDM-$Z_2$:
\begin{itemize}
    \item Only one $U(1)_H$ gauge symmetry is needed instead of two $Z_2$ symmetries.
    \item Like the SM, the N2HDM-$U(1)_H$ is built on local gauge symmetries and spontaneous symmetry breaking (i.e., unlike the N2HDM-$Z_2$ no soft-breaking is needed). 
    \item The N2HDM-$U(1)_H$ symmetry is more predictive than the N2HDM-$Z_2$ because it contains {one parameter less}.
\end{itemize}

If the $Z^\prime$ boson is decoupled from phenomenology, either because it is heavy or weakly interacting, the scalar sector of the N2HDM-$U(1)_H$ is close to the one of the N2HDM-$Z_2$, however, there are important differences:
\begin{itemize}
    \item In the N2HDM-$U(1)$ no $\lambda_5$ term is allowed, leading to $CP$ conservation. This feature is even conserved when the $Z^\prime$ is integrated out because of an automatic phase alignment {avoiding potentially dangerous effects in electric dipole moments.}.
    \item The $m_{12}^2$ term is absent before spontaneous symmetry breaking and induced by the VEV of $\phi$, either from the term $\mu H_1^\dag {H_2}\phi$ or ${\lambda_{\phi12}}(H_1^\dag {H_2})\phi^2$, depending on the charge assignment. Please note that the latter option was, to the best of our knowledge, not proposed before in the literature.
    \item While in N2HDM-$Z_2$, if $H$ is heavy, only symmetric decays into Higgs pairs; i.e., $H\to SS$ and $H\to hh$ are possible in the limit of zero mixing, in the N2HDM-$U(1)$ one expects naturally large branching ratios for $H\to S h$. Note that while in case (a), only asymmetric decays are unsuppressed, in case (b) also decays to identical scalars (e.g., $H\to SS$) can be sizable. 
\end{itemize}

The  {last difference} has important implications for the asymmetric $\approx 650\,$GeV excess in $b\bar b \gamma\gamma $. While even if $H$ is equipped with a sufficiently high production cross section (e.g., from direct top-quark Yukawa couplings of $H_1$), the N2HDM-$Z_2$ could not account for the preferred central value of the measurement as BR$(H\to S h)$ could not be sizable enough, taking into account the limits on the scalar mixing from Higgs coupling strength measurements at the LHC. However, the N2HDM-$U(1)$ can  {address}  this measurement, predicting signatures in $pp\to H(A)\to t\bar t$, $pp\to t\bar{t} H(A)\to 4t$ and $pp\to A\to SZ$, not far away from the current experimental limits.

Finally, let us point out that $Z$--$Z^\prime$ mixing, in general present in this model, can naturally account for the higher than expected value of the $W$ mass~\cite{Alguero:2022est}, as suggested by the measurement of the CDF-II Collaboration~\cite{Hays:2022qlw}. Together with the previous arguments this strongly motivates detailed studies of the N2HDM-$U(1)$ (including its Higgs sector) which, in our opinion, should be considered to be (at least) at the same level of interest as the standard N2HDM-$Z_2$ and therefore be examined with the {same scrutiny theoretical and experimental in the future, motivating more asymmetric di-Higgs searches at the LHC}. 
\medskip 

{\textit Note Added-} Recently, CMS presented updated results for low mass searches for new scalars decaying into $\gamma\gamma$~\cite{CMS-PAS-HIG-20-002}, confirming the previous excess.

\begin{acknowledgments}
We are very grateful to Michael Spira for useful comments on the manuscript and to Bruce Mellado for useful discussion. The work of A.\,C.~is supported by a professorship grant from the Swiss National Science Foundation (No.\ PP00P21\_76884). S.\,I. is supported by the Deutsche Forschungsgemeinschaft (DFG, German Research Foundation) under grant 396021762-TRR\,257. T.\,K.~is supported by the Grant-in-Aid for Early-Career Scientists from the Ministry of Education, Culture, Sports, Science, and Technology (MEXT), Japan, No.\,19K14706. This work is also supported by the Japan Society for the Promotion of Science (JSPS)  Core-to-Core Program, No.\,JPJSCCA20200002.
\end{acknowledgments}

\appendix
\section{Minimization and Mass Matrices \label{AppA}}

In this Appendix, we give the minimization conditions and mass matrices of both types, (a) and (b), of the N2HDM-$U(1)$. We define $H_1$, $H_2$ and $S$ as,
\begin{widetext}
\begin{align}
    H_1 = \begin{pmatrix}
  w^{+}_1\\[0.2cm]
  \frac{v_1 + \hat H + i \eta_1}{\sqrt{2}}
\end{pmatrix} \,,\qquad
H_2= \begin{pmatrix}
  w^{+}_2\\[0.2cm]
  \frac{v_2 + \hat h + i \eta_2}{\sqrt{2}}
\end{pmatrix}\,,  \qquad \phi = \frac{v_S + \hat S + i \eta_S}{\sqrt{2}} \,.
\end{align}
\subsection{Case (a): $|Q_H(\phi)|=|Q_H(H_1)-Q_H(H_2)|$}
The minimization conditions are
\begin{eqnarray}
   m^{2}_{11}+\frac{1}{2} \lambda_1 v^2_1 + \frac{1}{2} \lambda_{345} v^2_2 +\frac{1}{2} \lambda_{\phi 1} v^2_S + \mu v_S\frac{v_2 }{v_1} &=0 \,,  \nonumber \end{eqnarray}
\begin{eqnarray}
   m^{2}_{22}+\frac{1}{2} \lambda_2 v^2_2 + \frac{1}{2} \lambda_{345} v^2_1 +\frac{1}{2} \lambda_{\phi 2} v^2_S + \mu v_S\frac{ v_1 }{v_2} &=0\,,\\
   m^{2}_{S}+\frac{1}{2} \lambda_{\phi 1} v^2_1 + \frac{1}{2} \lambda_{\phi 2} v^2_2 +\frac{1}{2} \lambda_\phi v^2_S + \mu \frac{ v_1 v_2}{v_S} &=0\,, \nonumber 
\end{eqnarray}
where  $\lambda_{345}=\lambda_3 +\lambda_4 + \lambda_5^{\rm eff}$, and $\lambda_5^{\rm eff}\neq 0$ is only generated if the $Z^\prime$ is integrated out. The scalar squared-mass matrices are
\begin{align}
    M^2_\rho &= \begin{pmatrix}
  \lambda_1 v_1^2- \mu v_S\frac{v _2 }{v _1} & \lambda_{345} v_1 v_2+\mu v_S &
   \lambda_{\phi 1} v _1 v _S + \mu v_2 \\
 \lambda_{345} v_1 v_2+\mu v_S & \lambda _2 v _2^2-\mu v_S\frac{v _1 }{v _2} &
   \lambda_{\phi 2} v _2 v _S +\mu v_1\\
 \lambda_{\phi 1} v_1 v_S + \mu v_2 & \lambda_{\phi 2} v _2 v _S + \mu v_1 & \lambda_\phi v _S^2 - \mu \frac{v_1 v_2}{v_S} \\
\end{pmatrix}\,,  \\
M^2_\eta &= \begin{pmatrix}
 -\mu v_S \frac{ v_2 }{v _1} {-\lambda_5^{\rm eff} v_2^2}& \mu v _S {+\lambda_5^{\rm eff} v_1 v_2}& \mu v _2 \\
 \mu v _S {+\lambda_5^{\rm eff} v_1 v_2}& -\mu v_S \frac{v _1}{v _2} {-\lambda_5^{\rm eff} v_1^2}& -\mu v _1 \\
 \mu v_2 & -\mu v _1 & - \mu \frac{v_1 v_2}{v_S} \\
\end{pmatrix}\,,  \\
M^2_w &= \begin{pmatrix}
 - \mu v _S \frac{ v _2}{v _1}- (\lambda_4+\lambda_5^{\rm eff})\frac{ v _2^2}{2} & \mu v_S+ (\lambda_4+\lambda_5^{\rm eff})\frac{ v_1 v_2}{2} \\
 \mu v_S+ (\lambda_4+\lambda_5^{\rm eff})\frac{ v_1 v_2}{2} & - \mu v_S \frac{v _1 }{v _2}-(\lambda_4+\lambda_5^{\rm eff})\frac{ v_1^2}{2} \\
\end{pmatrix}\,,
\end{align}
which are defined via the bilinear potential terms
\begin{align}
V_{m^2} =&  \frac{1}{2}\begin{pmatrix}
  \hat H & \hat h & \hat S
\end{pmatrix} M^2_\rho  \begin{pmatrix}
  \hat H \\ \hat h \\ \hat S
\end{pmatrix}  + \frac{1}{2}\begin{pmatrix}
  \eta_1 & \eta_2 & \eta_S
\end{pmatrix} M^2_\eta  \begin{pmatrix}
  \eta_1 \\ \eta_2 \\ \eta_S
\end{pmatrix}  +  \begin{pmatrix}
  w^{-}_1 & w^{-}_2
\end{pmatrix} M^2_w  \begin{pmatrix}
  w^{+}_1 \\ w^{+}_2 
\end{pmatrix} \,.
\end{align}
The eigenvalues of the $CP$-odd and charged-Higgs masses are then given by
\begin{align}
    M_A^2 &= - \mu \left(  \frac{v_S v^2}{v_1 v_2} + \frac{ v_1 v_2}{v_S}\right) - \lambda_5^{\rm eff}v^2\,,\\
    M_{H^\pm}^2 &= - \frac{\mu v_S v^2}{v_1 v_2 } - \left(\lambda_4+\lambda_5^{\rm eff}\right)\frac{v^2 }{2}\,.
\end{align}

\subsection{Case (b): $|Q_H(\phi)|=|Q_H(H_1)-Q_H(H_2)|/2$}

The minimization conditions in this case are
\begin{align}
     m^{2}_{11}+\frac{1}{2} \lambda_1 v^2_1 + \frac{1}{2} \lambda_{345} v^2_2 +\frac{1}{2} \lambda_{\phi 1} v^2_S + \frac{1}{2} \frac{\lambda_{\phi 12}v_2 v^2_S}{v_1} &=0 \,,\nonumber \\
   m^{2}_{22}+\frac{1}{2} \lambda_2 v^2_2 + \frac{1}{2} \lambda_{345} v^2_1 +\frac{1}{2} \lambda_{\phi 2} v^2_S + \frac{1}{2} \frac{\lambda_{\phi 12}v_1 v^2_S}{v_2} &=0 \,,\nonumber \\
   m^{2}_{S}+\frac{1}{2} \lambda_{\phi 1} v^2_1 + \frac{1}{2} \lambda_{\phi 2} v^2_2 +\frac{1}{2} \lambda_\phi v^2_S + \lambda_{\phi 12} v_1 v_2 &=0\,,
\end{align}
and the squared-mass matrices are given by
\begin{align}
    M^2_\rho &= \begin{pmatrix}
  \lambda_1 v_1^2- \frac{\lambda_{\phi 12} v_2 v^2_S}{2 v_1} & \lambda_{345} v_1 v_2+ \frac{1}{2} \lambda_{\phi 12} v^2_S &
   \lambda_{\phi 1} v _1 v _S + \lambda_{\phi 12} v_2 v_S \\
 \lambda_{345} v_1 v_2+ \frac{1}{2} \lambda_{\phi 12} v^2_S & \lambda _2 v _2^2- \frac{\lambda_{\phi 12} v_1 v^2_S}{2 v_2} &
   \lambda_{\phi 2} v_2 v_S + \lambda_{\phi 12} v_1 v_S \\
 \lambda_{\phi 1} v _1 v _S + \lambda_{\phi 12} v_2 v_S & \lambda_{\phi 2} v _2 v _S + \lambda_{\phi 12} v_1 v_S & \lambda_\phi v_S^2  \\
\end{pmatrix}\,, \\
M^2_\eta &= \begin{pmatrix}
 -\frac{\lambda_{\phi 12} v _2 v _S^2}{2 v _1} - \lambda_5^{\rm eff} v_2^2 & \frac{1}{2} \lambda_{\phi 12} v _S^2 + \lambda_5^{\rm eff} v_1 v_2 & \lambda_{\phi 12} v _2 v _S \\
 \frac{1}{2} \lambda_{\phi 12} v _S^2 + \lambda_5^{\rm eff} v_1 v_2 & -\frac{\lambda_{\phi 12} v _1 v _S^2}{2 v _2} - \lambda_5^{\rm eff} v_1^2 & -\lambda_{\phi 12} v _1 v _S \\
 \lambda_{\phi 12} v _2 v _S & -\lambda_{\phi 12} v _1 v _S & -2 \lambda_{\phi 12} v _1 v _2 \\
\end{pmatrix}\,, 
 \\
M^2_w &= \frac{1}{2}  \begin{pmatrix}
-\frac{\lambda_{\phi 12} v _2 v _s^2}{v _1}  -(\lambda _4+\lambda_5^{\rm eff}) v _2^2 & \lambda_{\phi 12} v _s^2 + (\lambda _4+\lambda_5^{\rm eff}) v _1 v _2\\
 \lambda_{\phi 12} v _s^2 + (\lambda _4+\lambda_5^{\rm eff}) v _1 v _2 & -\frac{\lambda_{\phi 12} v _1 v _s^2}{v _2} -(\lambda _4+\lambda_5^{\rm eff}) v _1^2\\
\end{pmatrix}\,.
\end{align}
The eigenvalues of the $CP$-odd and charged-Higgs masses are then given by
\begin{align}
    M_A^2 & = - \lambda_{\phi 12}\left(  \frac{v_S^2 v^2}{2 v_1 v_2 } 
     + 2  v_1 v_2 \right) - \lambda_5^{\rm eff} v^2\,, \\
     M_{H^\pm}^2 & = - \frac{\lambda_{\phi 12} v_S^2 v^2}{2v_1 v_2}
     -  \left(\lambda _4+\lambda_5^{\rm eff}\right)\frac{v^2}{2} \,.
\end{align}
\end{widetext}

\bibliographystyle{utphys}
\bibliography{references}

\end{document}